\documentstyle[psfig,12pt]{article}
\def\cm{\,{\rm cm}}
\def\km{\,{\rm km}}

\def\mpc{\,{\rm Mpc}}

\def\ev{\,{\rm eV}}

\def\gev{\,{\rm GeV}}

\def\sr{\,{\rm sr}}
\def\km{\,{\rm km}}
\def\sec{\,{\rm sec}}
\def\yr{\,{\rm yr}}

\def\la{\mathrel{\mathpalette\fun <}}
\def\ga{\mathrel{\mathpalette\fun >}}
\def\fun#1#2{\lower3.6pt\vbox{\baselineskip0pt\lineskip.9pt
  \ialign{$\mathsurround=0pt#1\hfil##\hfil$\crcr#2\crcr\sim\crcr}}}
\newcommand{\beq}{\begin{equation}}
\newcommand{\eeq}{\end{equation}}

\begin{document}
\vspace{-0.8in}{\flushright{\hfill astro-ph/9803029}}
\begin{center}
{\large \bf Ultrahigh Energy Cosmic Rays from Topological Defects
--- Cosmic Strings, Monopoles, Necklaces, and All That}\footnote{Invited
talk given at the Workshop on ``Observing the Highest Energy Particles
($> 10^{20}\ev$) from Space'', College Park, Maryland, USA, November 13 --
15, 1997, to be published in the Proceedings (AIP).} 

\vspace{.2in}

Pijushpani Bhattacharjee\footnote{NAS/NRC Resident Senior Research
Associate at NASA/GSFC on sabbatical leave from Indian Institute of
Astrophysics, Bangalore-560 034. India.}\\

\vspace{0.1in}

{\it Laboratory for High Energy Astrophysics,\\
NASA/Goddard Space Flight Center, Code 661, \\
Greenbelt, MD 20771. USA}.\\

\vspace{0.1in}

{\it and}

\vspace{0.1in}

{\it Indian Institute of Astrophysics\\
Koramangala, Bangalore 560 034, INDIA}\\

\end{center}

\vspace{0.3in}

\centerline{\bf Abstract}
\medskip
The topological defect scenario of origin of the observed highest energy
cosmic rays is reviewed. Under a variety of circumstances,
topological defects formed in the early Universe can be sources of  
very massive particles in the Universe today. The decay products of
these massive particles
may be responsible for the observed highest energy cosmic ray particles
above $10^{20}\ev$. Some massive particle production processes
involving
cosmic strings and magnetic monopoles are discussed. We also discuss the
implications of results of certain recent numerical
simulations of evolution of cosmic strings. These results (which remain to
be confirmed by independent simulations)   
seem to show that massive particle production may be a generic feature of
cosmic strings, which would make cosmic strings an inevitable source of
extremely high energy cosmic rays with potentially detectable flux. At
the same time, cosmic strings are severely constrained
by the observed cosmic ray flux above $10^{20}\ev$, if massive particle
radiation is the dominant energy loss mechanism for cosmic strings.  
\section*{Introduction}
Cosmic Topological Defects (TDs)\cite{kibble,td} such as magnetic
monopoles,
cosmic strings, domain walls, and various hybrid TD systems
consisting of these basic kinds of TDs, are predicted to form in the early
Universe as a result of symmetry-breaking phase transitions envisaged in
unified theories of elementary particle interactions. Under a variety of
circumstances these TDs 
can be sources of extremely massive unstable particles in the universe 
today\cite{bhs,hill,hsw,cusp,pbkofu,br,gk,bs,necklace}. The masses
$m_X$ of these so-called ``X'' particles
(the quanta of the massive gauge- and higgs fields of the
underlying spontaneously broken gauge theory) would typically
be of order the symmetry-breaking energy scale at which the
relevant TDs were formed, which, in Grand Unified Theories (GUTs), can be
as large as $\sim 10^{16}\gev$. The decay of these X particles can give
rise to extremely energetic photons, neutrinos and nucleons with energy up
to $\sim m_X$. If the X particle production rate from TDs is large enough, 
these extremely energetic particles may be detectable by ground-based 
as well as space-based 
large-area detectors being planned for detecting ultrahigh
energy (UHE) (i.e., energy $\ga10^{9}\gev$) cosmic rays. These cosmic ray
detectors may thus provide us with a
tool for studying the signature of TDs and thus of GUT scale
physics. 

There is currently much interest in the possibility that the   
Extremely High Energy (EHE) Cosmic Ray (EHECR) events --- those  
with energies above $10^{11}\gev$ reported recently\cite{hecrevents} --- 
may be due to decays of massive X particles originating from
TDs\cite{ssb,bs,slsb,slc,slsc,ps,necklace}. 
This possibility is of interest in view of the fact that the
energies associated with the EHECR events are hard to
obtain\cite{hillas,ssb,norman} within the 
standard diffusive shock acceleration mechanism\cite{drury} that involves 
first-order Fermi acceleration of charged particles at relativistic shocks
associated with known powerful astrophysical objects; see, however,
Ref.~\cite{biermann}. 
In addition, there is the problem of
absence of any obviously identifiable sources for these EHECR
events\cite{es,ssb}. 
These problems are avoided in the TD scenario in a natural way. Firstly,
no acceleration mechanism is needed: The decay products of the X particles
have energies up to $\sim m_X$ which can be as large as, say,
$10^{16}\gev$. 
Secondly, the absence of obviously identifiable sources is not a
problem because TDs need not necessarily be associated with any
visible or otherwise
active astrophysical objects such as AGNs or radio galaxies.   

The basic ideas of the TD scenario of origin of EHECR have been reviewed
in a number of discussions in the past; see, e.g., Refs.
\cite{pbkofu,pbicnapp,pbtokyo,siglheidel}. 
Detailed calculations of the predicted spectra of nucleons, photons, and
neutrinos in the TD scenario have been done in the past several
years\cite{bhs,abs,bs,slc,slsb,ps,slsc,yoshida}. Constraints on the TD
scenario imposed by experimental data on EHECR and diffuse 
gamma ray background 
have also been discussed\cite{chi,slc,ps,slsc}. 

In what follows, I first discuss briefly some of the basic aspects of
topological defects and their formation in the early Universe. 
The basic steps in the calculation of the `observable' particle spectra
resulting from the decay of the X particles is discussed.  
A simple benchmark calculation of the X particle production rate 
required to explain the observed EHECR particle flux is performed. I then
discuss three specific X particle production mechanisms involving (1)
``ordinary'' cosmic strings, (2) monopolonia --- metastable
monopole-antimonopole bound states, and (3) cosmic ``necklace'' --- a
system of monopoles on strings. 
I also discuss the implications of the results of certain recent
numerical simulations
of evolution of cosmic strings in the Universe. These results, if
confirmed by independent simulations, would imply that massive 
X-particle production and hence cosmic ray production might be a
generic feature of cosmic strings, which would make cosmic strings an
inevitable source of EHECR with potentially detectable flux. 
Indeed, in this case, we shall see that the measured EHECR flux already
puts severe 
constraint on the energy scale of symmetry-breaking associated
with any cosmic string forming phase transition in the early
Universe. 

I use natural units with $\hbar=c=k_B=1$ throughout, unless otherwise
stated. 

\section*{Topological Defects and X Particle Production: A
Brief Review}
Topological Defects are sometimes characterized as ``exotic''. 
In actual fact, TDs are routinely seen, measured, and studied in condensed
matter systems in laboratories. TDs form during phase transitions associated
with the phenomenon of spontaneous symmetry breaking (SSB), which is a
central concept in condensed matter physics as well as in the 
Standard Model of particle physics. 
Well-known examples of TDs in condensed matter systems are vortex
lines in superfluid helium, magnetic flux tubes in type-II
superconductors, disclination lines and `hedgehogs' in nematic liquid
crystals, and so on. Perhaps what is perceived as exotic is the existence
of TDs in the cosmological context. However, it has been realized for
quite some time now, particularly since the early seventies, that
our Universe in its early stages must have behaved very much like
condensed matter systems. Indeed, ideas of
unified gauge theories of elementary particle interactions taken together
with the hot big-bang model of the early Universe necessarily imply that
our Universe in its early history has passed through a sequence of
symmetry-breaking phase transitions as it expanded and cooled through
certain critical temperatures. Depending on the symmetry breaking pattern,
one or more kinds of the three basic kinds of topological defects --- 
magnetic monopoles, cosmic strings and domain walls --- could be formed
during some of these phase transitions\cite{kibble,td}. In fact, 
formation of magnetic monopoles, is {\it inevitable} in practically all 
Grand Unified Theories (GUTs) that provide a unified description of the
electroweak and strong interactions\footnote{The inevitability of monopole
formation leads to the well-known ``monopole problem'' of cosmology, which
historically was one of the ``problems'' that motivated the idea of {\it
inflationary cosmology}; for a review, see Ref.\cite{kt}}. The monopoles
are analogous to the `hedgehogs' in nematic liquid crystals and appear
whenever the unbroken symmetry group possesses a local U(1) symmetry. 
The `global' cosmic strings, which arise in breaking of global U(1)
symmetry, are similar to vortex
filaments in superfluid helium, and the `local' or `gauge' cosmic strings
arising from breaking of a local U(1) symmetry are similar to the magnetic
flux tubes in type-II superconductors. Cosmic domain walls appear whenever
a discrete symmetry is spontaneously broken. 

Interestingly, it has
recently become possible to simulate the analogue of cosmic string
formation in the early Universe by means of 
laboratory experiments\cite{helium3} on vortex-filament formation in the
superfluid
transition of $^3{\rm He}$, which occurs at a temperature of a few
millikelvin. The results of these experiments have provided
striking confirmation of the basic Kibble-Zurek
picture\cite{kibble,td,zurek} of topological defect formation in general,
which was initially developed within the context of defect formation in       
the early Universe\cite{kibble}. 
\subsection*{Symmetry Breaking Phase Transitions and
Formation of Topological Defects}
During a symmetry breaking phase transition,  
the system under consideration undergoes a transition from a
state of higher symmetry to one of a lower (reduced) symmetry at a
critical temperature during the cooling of the system. In 
spontaneous symmetry breaking (SSB), the system
below the critical temperature possesses multiple degenerate ground states 
rather than a unique ground state. 
These degenerate ground states differ from each other by the `phase angle'
or some
internal degrees of freedom of the ``order parameter'' field (OPF) whose
absolute magnitude (which is same for all the ground states) is
a measure of the order (or lack of symmetry) in the 
system. (In particle physics, the OPF is the ``higgs'' field.)  
The symmetry under consideration is invariance
of the energy (or more precisely the Lagrangian for the OPF or the higgs
field) under transformations that change the 
`phase' of the OPF. The Lagrangian is always invariant under these
transformations because, by construction, it depends only on the absolute
magnitude of the OPF and not on its `phase'. However,
the ground states transform among each other under the action of these
transformations, and so any chosen ground state is clearly not
invariant under the transformations --- the symmetry is
spontaneously broken. The existence of multiple degenerate ground
states is the defining characteristic of the phenomenon of spontaneous 
symmetry breaking. By convention, the absolute value of the OPF 
is taken to be zero in the high-temperature unbroken-symmetry
phase and unity in the low-temperature broken symmetry phase. 

Because of the availability of these multiple degenerate ground
states in
the low temperature phase, different parts of the system may choose to
settle down to different ground states when making the transition to the
low temperature phase, especially so if the transition happens in an
out-of-equilibrium situation. Indeed, one expects that the choice of the
phase of the OPF in regions separated by more than the correlation length
of the thermal fluctuations of the OPF will be uncorrelated. The
correlation length will always be finite in a finite physical system.   
In the context of the Universe as a whole, there is also an upper limit to
the correlation length at any time $t$, namely, the causal horizon length
$\sim ct$. Thus the choice of the phase of the OPF will be random and in
general different in different parts
of the Universe separated by more than the causal horizon length at the
time of phase transition. This often leads to `obstruction' in the way
of uniform completion of the transition throughout the bulk of the system.
Indeed, it is often the case that the random choice of different ground
states in different regions leads to some
regions being forced to remain in the unbroken symmetry phase. These
regions are the `topological defects'. 

The topological nature of
the defects becomes clear when one considers the configuration of the OPF
in the low temperature phase at the end of the phase transition.   
The choice of the `phase' of the OPF 
corresponding to different ground states in different parts of the system
may be such that, in order to avoid
energetically unfavored discontinuity in the spatial variation of the OPF,
the magnitude of the OPF is forced to be zero on some
geometrical points, lines, or surfaces, which define the `center' of the 
defects. Actually, a defect has a finite size dictated by the need 
to minimize the overall energy of the system; the absolute value of
the OPF increases {\it gradually} from zero at the center to its
broken-symmetry-phase value at a finite distance from the center. 

The topological stability of the defects is due to non-trivial
topological `winding' of the OPF configuration around the defect center. 
For example, in the case of the linear defect (the vortex filament in the 
superfluid helium, for example), the OPF (the wave function 
of the condensed helium atoms, for example) is a complex number whose 
phase turns by an integral multiple of 
$2\pi$ as one makes a complete circuit along a closed curve around any
point on the defect line (the closed curve being on the normal plane
cutting the defect line at the given point). Once formed, such a
configuration cannot `unwind' by itself and is thus topologically stable. 

In the context of spontaneously broken gauge theories, explicit 
analytical and/or numerical finite-energy,
extended, topologically stable solutions for the higgs- and gauge
field configurations representing cosmic strings, monopoles and domain
walls are known; see Ref.\cite{td} for review. 
In the broken-symmetry phase, a higgs field 
responsible for spontaneous symmetry breaking of a local gauge theory  
is massive, as are the gauge bosons of the theory, the mass scale being
set by the absolute magnitude of the higgs field (more precisely,
its vacuum expectation value) in the broken symmetry phase. The size of
the `core' of a defect is of order $m_X^{-1}$, where $X$ represents the
higgs or the gauge field. Within the core, the symmetry remains unbroken,
and the energy density (associated with the higgs and the gauge fields) 
is higher within the core than outside. The topological stability of the
defect ensures the `trapping' of the excess energy within the core of the
defect, which is what makes a defect massive. It can be shown that the
mass scale of a defect is fixed by the energy (or temperature) scale of
the symmetry breaking phase transition at which the defect is formed.
Thus, if we denote by $T_c$ the critical temperature of the defect-forming
phase transition in the early Universe, then the mass of a monopole formed
at that phase transition is roughly of order $T_c$, 
the mass per unit length of a cosmic string is of order $T_c^2$, and the
mass per unit area of a domain wall is of order $T_c^3$. 

The X particles are the quanta of excitations of the higgs and gauge
fields. In the broken symmetry phase, these quanta are massive, their mass
is also roughly of order $T_c$. These massive quanta typically have very
short life times, and so they all decayed away quickly soon after the
phase transition in the early Universe, and none of those X particles
survive in the present Universe. This is to say that outside of a
defect in the low-energy Universe today, the higgs and the gauge fields
are in their ground states and no excitations of the X particle quanta are
present. However, inside the core of a defect, the
symmetry is unbroken --- the X particles are massless inside. Topological
stability of the defect prevents this ``piece of the early Universe''
trapped inside the defect from decaying. If, however, there is a process
which removes this topological protection, then the energy trapped inside
the defect will dissociate into quanta of the X particles which, being now
in the broken-symmetry phase, will be massive and short lived, decaying
into elementary particles such as quarks and leptons which, in turn, would
eventually materialize into energetic nucleons, photons, neutrinos, etc., 
that might contribute to the observed EHECR flux. 

Production of X particles from TDs may happen in a variety of ways
directly or indirectly related to local removal of topological stability
of (parts of) TDs. Examples include `cusp' evaporation from cosmic
strings\cite{cusp}, collapse of
macroscopic cosmic string loops\cite{br}, shrinking of cosmic string loops
to radii of order $m_X^{-1}$ \cite{gk}, annihilation of a 
monopole with an antimonopole\cite{hill,bs}, and so on. 
In the case of current-carrying superconducting cosmic
strings\cite{witten,hsw}, the charge carriers (which could be quanta of a
superheavy fermion field trapped in `zero mode' inside the string, or a
charged scalar field living inside the string due to energetic reasons) 
are expelled from the string when the current
on the string reaches a critical value --- in the vacuum away from the
string, these charge carriers act as the massive X particles. In the case
of `ordinary' cosmic strings, recent field theory
simulations~\cite{vincent} of
evolution of cosmic strings in the early Universe show that a cosmic
string network loses energy directly into oscillations of the underlying
gauge and higgs fields `constituting' the string, which in quantum theory,
correspond to quanta of massive X particles --- a result which has
important implications for both cosmic strings as well as for EHECR; this
result, however, remains to be confirmed by independent simulations.  
\subsection*{X-Particle Production Rate}
The number density of X particles produced by TDs per unit time,
$dn_X/dt$, can be generally written as\cite{bhs}
\beq
{dn_X\over dt}(t)= {Q_0\over m_X} \left({t\over
t_0}\right)^{-4+p}\,,\label{xrate}
\eeq
where $t_0$ denotes the present epoch, and $Q_0$ is the rate of energy
density injected in the form of X particles in the present epoch. The
quantity $Q_0$ and the
parameter $p$ depend on the specific TD process under consideration. In
writing Eq.~(\ref{xrate}) it is assumed that the only time scale in the
problem is the hubble time $t$ and that any other time scale involved can
be expressed in terms of the hubble time. Similarly, we assume that any
energy scale involved in the problem is expressible in terms of the energy
scale $\eta$ of the symmetry breaking at which the TDs under consideration
are formed. (Note that $m_X$ is fixed by $\eta$.) These assumptions are
sometimes expressed by saying that the TDs under consideration evolve
in a scale independent way. Indeed, it turns out that Eq.~(\ref{xrate})
is a phenomenologically good parametrization for the specific TD processes
studied so far. For example, $p=1$ for a process of X particle production
from cosmic string loops~\cite{br}, for a process involving collapsing
monopole-antimonopole bound states~\cite{hill,bs}, for a process of
particle production from monopole-string systems called ``necklaces''
~\cite{necklace}, and so on, while $p=0$ for a process involving
superconducting cosmic string loops~\cite{hsw}. 
 
Note that X particle production from TDs may occur continually at all 
epochs after the formation of the relevant TDs in the early Universe.
However, only those X
particles produced in the relatively recent epochs and at relatively
close-by, non-cosmological distances ($\la100\mpc$) are relevant for the
question of EHECR. This is because, nucleons of energies above
$10^{11}\gev$ produced by the decay of X particles occurring at distances
much larger than $\sim 50\mpc$ suffer drastic energy loss during their
propagation, due to photopion production on the the cosmic microwave
background radiation (CMBR) fields (the so-called ``GZK
effect''~\cite{gzk}), and hence do not survive as EHECR
particles today. Distance of sources of photons of energies above
$\sim10^{11}\gev$ are also similarly restricted  
due to absorption through $e^+e^-$ production on the radio background
photons (see, e.g., ~\cite{abs}). The neutrinos, however, can survive 
from much earlier cosmological epochs, and may, if detected, prove to be
the ultimate discriminant between a TD scenario and a more conventional
scenario of origin of EHECR.   

\section*{From X Particles to `Observable' Particles}
The X particles released from TDs would decay typically into quarks and
leptons. The life-time $\tau$ is typically $\sim(\alpha m_X)^{-1}$ (where
$\alpha\sim {\rm few}\times 10^{-2}$),
which for $m_X\sim10^{16}\gev$ is $\sim10^{-39}\sec$ or so. The decay is,
therefore, essentially instantaneous at late cosmological epochs of
interest to us. The quarks would hadronize (typically on a strong
interaction time scale $\sim10^{-23}\sec$, i.e., again practically
instantaneously) by producing jets of hadrons, most of which would
eventually be light mesons (pions) with a small admixture (typically
$\la10\%$) of baryons and antibaryons (nucleons and antinucleons). The
neutral pions decay to two photons, while the decay of charged pions 
gives rise to neutrinos. Some leptons (charged as well as neutral) could
also be produced directly from the X particle decay. But by far the
largest number of nucleons, photons, and neutrinos, would be produced
through the hadronic channel. The spectra of produced particles are,
therefore, essentially determined by the process of fragmentation
of quarks/gluons into hadrons as described by QCD. 
\subsection*{Quark $\to$ Hadron Fragmentation Spectrum}
The exact process by which a single high
energy quark gives rise to a jet of hadrons is not known; it involves some
kind of non-perturbative physics that is not well
understood. However, different semi-phenomenological approaches have
been developed which describe the hadronic ``fragmentation spectrum'' of
quarks/gluons that are in good agreement with the currently
available experimental data on
inclusive hadron spectra in quark/gluon jets in a variety of high
energy processes. 

In these approaches, the process of production of a jet containing a large
number of hadrons, by a single high energy quark (or gluon),  
is `factorized' into two stages.  
The first stage involves `hard' processes involving large momentum 
transfers, whereby the initial high energy
quark emits `bremsstrahlung' gluons which themselves create more 
quarks and gluons through various QCD processes. These hard processes are
well described by perturbative QCD. Thus a single high energy quark gives
rise to a `parton cascade' --- a shower of quarks and gluons --- which,
due to the high energy nature of the process, is confined in a narrow cone
or jet whose axis lies along the direction of propagation of the
original quark. This first stage of 
the process, i.e., the parton cascade development, described by
perturbative QCD, is terminated at a cut-off value, $\langle
k_\bot^2\rangle^{1/2}_{\rm cut-off}\sim 1\gev$, of the typical transverse
momentum. Thereafter, the second stage, involving 
the non-perturbative ``confinement'' process, takes over binding the
quarks and gluons into color neutral hadrons. This second stage is usually
described by one of the available phenomenological hadronization models 
such as the LUND string fragmentation
model~\cite{lund} or the cluster fragmentation model~\cite{cluster}. 
Detailed Monte Carlo numerical codes now
exist~\cite{cluster,jetset,ariadne} which incorporate the two stages
outlined above. These codes provide a reasonably good description of a
variety of relevant experimental 
data. Clearly, however, this is a numerically intensive approach. 
\subsubsection*{Local Parton-Hadron Duality}  
There is an alternative approach that is essentially analytical 
and has proved very fruitful in terms of its ability to
describe the gross features of hadronic jet systems, such
as the inclusive spectra of particles, the particle multiplicities and
their correlations, etc., reasonably well. This approach is based on the
concept of ``Local Parton Hadron Duality'' (LPHD)~\cite{lphd}. Basically,
in this approach, the second stage involving the non-perturbative
hadronization process mentioned above is ignored, and the hadron spectrum
is taken to be the same, up to an overall normalization constant, as
the spectrum of partons (i.e., quarks and gluons) in the parton
cascade after evolving the latter all the way
down to a cut-off transverse momentum $\langle
k_\bot^2\rangle^{1/2}_{\rm cut-off}\sim R^{-1}\sim$ few hundred MeV,
where $R$ is a typical hadronic size. At present the only justification
for such an approach seems to be that it gives a remarkably good
description of the experimental data including recent experimental results
from LEP, HERA and TEVATRON ~\cite{lphdreview}. Justification of LPHD at
a more fundamental theoretical level, however, is not yet available.
Nevertheless, it serves as a good phenomenological tool.  

The main assumption 
in LPHD is that the actual hadronization process, i.e., the
conversion of the quarks and gluons in the parton cascade into color
neutral hadrons, occurs at a low virtuality scale of order of a
typical hadron mass independently of the energy of the cascade initiating 
primary quark, and involves only low momentum transfers and 
local color `re-arrangement' which somehow does not drastically alter the
form of the momentum spectrum of the particles in the parton cascade
already determined by the `hard' (i.e., large momentum transfer)
perturbative QCD processes. Thus, the non-perturbative hadronization
effects are lumped together in an `unimportant' overall normalization
constant which can be determined phenomenologically.  

A good quantitative description of the perturbative QCD stage of the
parton cascade evolution is provided by 
the so-called Modified Leading Logarithmic
Approximation (MLLA)~\cite{mlla} of QCD, which allows the parton
energy spectrum to be expressed analytically in terms of 
functions depending on two free parameters, namely, the
effective QCD scale $\Lambda_{\rm eff}$ (which fixes the effective
running QCD coupling strength $\alpha_s^{\rm eff}(\widetilde{Q}^2)$) and
the
transverse momentum cut-off $\widetilde{Q}_0$.
For the case $\widetilde{Q}_0=\Lambda_{\rm eff}$, the analytical result
simplifies
considerably, and one gets what is referred to as the ``limiting
spectrum''~\cite{lphd,lphdreview}. For asymptotically high energies of
interest, i.e., for 
energies $E_{\rm jet}$ of the original jet-initiating quark satisfying 
$E_{\rm jet}\gg \Lambda_{\rm eff}$, the
limiting spectrum can be approximated by a Gaussian in
the variable $\xi\equiv\ln(1/x)$, with $x\equiv E_{\rm parton}/E_{\rm
jet}$,
$E_{\rm parton}$ being the energy of a quark (parton) in the jet: 
\beq 
x\frac{dN_{\rm parton}(Y,x)}{dx}\approx\frac{N_{\rm
parton}(Y)}{\sigma\sqrt{2\pi}}
\exp\left[-\frac{(\xi-\bar{\xi})^2}{2\sigma^2}\right]\,,\label{gauss}
\eeq
where $Y\equiv\ln\left(E_{\rm jet}/\Lambda_{\rm eff}\right)$, $\,\,\,
\bar{\xi}\approx Y/2$, $\,\,\,2\sigma^2=\left(\frac{bY^3}{36
N_c}\right)^{1/2}\,\,$ with 
$b\equiv (11 N_c - 2 N_f)/3\,\,\,$, $N_c$ being the number of colors and
$N_f$
the number of flavors of quarks involved, and
$N_{\rm parton}(Y)\sim\exp\{(16N_cY/b)^{1/2}\}$ is the average total
multiplicity of the partons in the jet. 

Eq.~(\ref{gauss}) gives us the spectrum of the partons in
the jet. By LPHD hypothesis, the shape of the hadron spectrum, 
$dN_h/dx$ (with $x=E_h/E_{\rm jet}\,,\,\,$ $E_h$ being the energy of a
hadron in the jet), is given by
the same form as in Eq.~(\ref{gauss}), except for an overall
normalization constant  
%
%
that takes account of the effect of conversion of partons into hadrons. 
Phenomenologically, for given values of
$\Lambda_{\rm eff}$ and $E_{\rm jet}$, the normalization constant
can be determined simply from overall energy conservation,
i.e., from the condition $\int_0^1 x\frac{dN_h(Y,x)}{dx}\, dx = 1\,$.
The value of 
$\Lambda_{\rm eff}$ is not known {\it a priori}, but a fit to the 
inclusive charged particle spectrum
in $e^+e^-$ collisions at center-of-mass energy $E_{\rm cm}=2 E_{\rm
jet}\sim 90\gev$ (Z-resonance) gives $\Lambda_{\rm eff}^{\rm ch}\sim$ 250
MeV.  

Note that, within the LPHD picture, there is no way of distinguishing
between various different species of hadrons. Phenomenologically, the
experimental data can be fitted by using different values of $\Lambda_{\rm
eff}$ for different species of particles depending on their masses. For
our consideration of
particles at EHECR energies, all particles under consideration will be
extremely relativistic, and since, in our case, $E_{\rm jet}\sim m_X/2\gg
\Lambda_{\rm eff}$, the hadron spectrum will be relatively insensitive to
the exact value of $\Lambda_{\rm eff}$. Also, one can safely assume that
at the asymptotically high energies of our interest, all hadrons ---
mesons as well as baryons --- will have the same spectrum. However, the
dominant species of particles in terms of their overall number will be the
light mesons (pions); baryons typically constitute a fraction of $\la$ (3
-- 10)\% 
as indicated by existing collider data. For more details on
various phenomenological aspects of the LPHD hypothesis, see the
reviews~\cite{lphdreview}. 
\subsection*{Nucleon, Photon and Neutrino Injection Spectra}
Using the MLLA + LPHD hadron spectrum discussed above, and assuming that
each X particle on average undergoes $N$-body decay 
(typically $N\leq 3$) to $N_q$ quarks (including 
antiquarks) and $N_\ell$ leptons (neutrinos and/or charged leptons), so 
that $N=N_q + N_\ell$, and assuming that the energy $m_X$ is shared
roughly equally by the $N$ primary decay products of the $X$, the
nucleon injection spectrum, $\Phi_{\rm N}(E_i,t_i)$, from the decay of all
X particles from TDs at any time $t_i$ can be written as 
\beq
\Phi_{\rm N}(E_i,t_i)=\frac{dn_X(t_i)}{dt_i}N_q f_N {1\over E_i}N_{\rm
norm} 
{1\over \sigma\sqrt{2\pi}}\exp\left[-\ln^2 ({x_*\over
x})\right]\,,\label{N_inj}
\eeq
where $E_i$ denotes the energy at injection, ${dn_X\over dt_i}$ is the
number of X particle released per unit
volume per unit time at time $t_i$, $f_{\rm N}$ is the nucleon fraction in
the hadronic jet produced by a single quark, 
$x=N E_i/m_X$, $x_*=(N \Lambda_{\rm eff}/m_X)^{1/2}$, and $N_{\rm norm}$
is the normalization constant defined by 
\beq
N_{\rm norm}(m_X)=\left(\int_0^1\, dx\, {1\over \sigma\sqrt{2\pi}}
\exp\left[-\ln^2 ({x_*\over x})\right]\right)^{-1}\,.\label{norm}
\eeq

An important point about the nucleon injection spectrum given by
Eq.~(\ref{N_inj}) is that, unlike the spectrum predicted in the standard
diffusive shock acceleration theory (see, e.g., 
Refs.~\cite{drury,hillas,norman}), the injection spectrum in the TD
scenario (or, for that matter, in any non-acceleration scenario in which
the energetic particles arise from decay of massive elementary particles),
is not, in general, a power-law in energy. Although, in the energy regions
of our interest, the spectrum (\ref{N_inj}) can be
approximated~\cite{pbkofu} by power-law segments ($\propto
E_i^{-\alpha}$), the power-law index $\alpha$, in the energy regions of
interest, is generally smaller than that in shock acceleration theories
--- the latter typically predict $\alpha\geq 2$. In other words, the
injection spectrum in non-acceleration theories is generally harder (or
flatter) compared to that in conventional acceleration theories. This fact
has important consequences; it leads to the prediction of a pronounced
``recovery''~\cite{bhs} of the evolved nucleon spectrum after a partial
GZK ``cut-off'' and the consequent flattening of the spectrum above
$\sim10^{11}\gev$. A relatively hard spectrum may also naturally give rise
to a ``gap'' in the measured EHECR spectrum~\cite{slsb}.  

The photon injection spectrum from the decay of the neutral pions ($\pi^0
\to 2\gamma$) in the jets is given by 
\beq
\Phi_\gamma(E_i,t_i)\simeq 2\int_{E_i}^{m_X/N} {dE\over E}
\Phi_{\pi^0}(E)\,,\label{photon_inj}
\eeq
where $\Phi_{\pi^0}\simeq {1\over3} {1-f_N\over f_N} \Phi_N$ is the
neutral pion spectrum in the jet. 

Similarly, the neutrino ($\nu_\mu+\bar{\nu}_\mu$) injection spectrum
resulting from the charged pion decay [$\pi^\pm \to \mu^\pm\nu_\mu
(\bar{\nu}_\mu)$] can be written as ~\cite{stecker_nu,bhs} 
\beq
\Phi_(\nu_\mu+\bar{\nu}_\mu)(E_i)\simeq 2.34\int_{2.34E_i}^{m_X/N}{dE\over
E} \Phi_{(\pi^++\pi^-)}(E)\,,\label{nu_inj}
\eeq
where $\Phi_{(\pi^++\pi^-)}\simeq {2\over3} {1-f_N\over f_N} \Phi_N$. 

The decay of each muon (from the decay of a charged pion) produces two
more neutrinos and an electron (or positron): $\mu^\pm\to e^\pm
\nu_e(\bar{\nu}_e)\, \bar{\nu}_\mu (\nu_\mu)$. Thus each charged pion
eventually gives rise to three neutrinos: one $\nu_\mu$, one
$\bar{\nu}_\mu$ and one $\nu_e$ (or $\bar{\nu}_e$), all of roughly the
same energy. So the total $\nu_\mu + \bar{\nu}_\mu$ injection spectrum 
will be roughly twice the spectrum given in Eq.~(\ref{nu_inj}), while the
total $\nu_e + \bar{\nu}_e$ spectrum will be roughly same as that in
Eq.~(\ref{nu_inj}). 

Note that, if the hadron spectrum in the jet is generally approximated by
a power-law in energy, then nucleon, photon and neutrino injection spectra 
will also have the same power-law form all with the same power-law index. 

It is worth emphasizing here that while using the LPHD hadron spectrum in
the analysis of the TD scenario of
EHECR one should keep in mind that there is a great deal of uncertainty
involved in extrapolating the QCD based hadron spectra --- which
have been tested so far only at relatively 
`low' energies of $\sim$ 100 GeV --- to the extremely high energies of
$\sim10^{15}\gev$ or so.
\subsection*{Evolution of the Particle Spectra}
\subsubsection*{Nucleons}
The evolution of the nucleon spectrum is mainly governed by interactions
of
the nucleons with the CMBR. The relevant interactions are pair production
by protons ($p\gamma_b\to p e^+e^-$), photoproduction of single or
multiple pions by nucleons $N$ ($N\gamma_b \to N n \pi, \, n\geq1$), and
neutron decay. Here $\gamma_b$ stands for a background photon, in this
case, a CMBR photon. At EHECR energies, the photopion production is the
dominant process. This is a drastic energy loss process for nucleons, and
is the basis of the well-known prediction of the GZK cut-off~\cite{gzk} of 
the evolved EHECR nucleon spectrum at energies above $\sim 10^{11}\gev$.   
This process also limits the distance of a possible source of the observed
EHECR particles to distances less than $\sim 50\mpc$ \cite{gzk,ssb,es}. 
\subsubsection*{$\gamma$-rays}
The $\gamma$-rays at EHECR energies interact via pair production (PP:
$\gamma\gamma_b \to e^+e^-$) and double pair production 
(DPP: $\gamma\gamma_b \to e^+e^-e^+e^-$), while the electrons (positrons)
interact via inverse Compton scattering (ICS: $e\gamma_b \to e^\prime
\gamma$) and triplet pair production (TPP: $e\gamma_b \to e e^+ e^-$). In
addition, the electrons (positrons) suffer synchrotron energy loss in the
extragalactic magnetic field (EGMF). The background photons ($\gamma_b$)
involved in the PP process are mainly the universal radio background (URB) 
photons for $\gamma$-rays above $\sim 10^{19}\ev$, the CMBR photons for
$\gamma$-rays between $\sim 10^{14}\ev$ and $\sim10^{19}\ev$, and the
infrared and optical (IR/O) background photons for $\gamma$-rays below
$\sim10^{14}\ev$. 

The evolution
of the $\gamma$-ray spectrum is complicated due to the fact that PP and 
ICS processes
together lead to development of electromagnetic (EM) cascades, whereby 
any electromagnetic energy in the form of $\gamma$-rays and/or 
electrons (positrons) released at an energy which is above the
threshold for PP process on photons of a particular background,
cascades down to progressively lower energies through a cycle of PP
interactions (on the photons of progressively higher energy
backgrounds) and ICS interactions (mainly on the CMBR). 
The cascading has the overall effect of increasing
the $\gamma$-ray flux at EHECR energies because it causes an effective
increase of the attenuation length of these $\gamma$-rays. 
Further cascading
by any cascade photon stops either if the remaining path length of
the cascade photon, as it propagates from its point of creation to the
observation point, becomes less than the mean free path for the
PP process on the relevant background photons, or if the
energy of the propagating cascade photon falls below the threshold for
the PP process. 
Thus, depending on the distance at which they are first
injected, $\gamma$-rays of EHECR energies can, after propagation,  
give rise to a $\gamma$-ray spectrum
that may span the energy range from (say) a few tens of MeV all the way
up to EHECR energies. This, of course, means that any model of
electromagnetic energy injection at EHECR energies has to
meet the constraint that the resulting cascade $\gamma$-ray flux at
lower energies should not exceed the measured flux at those energies. 
In the context of TD models of EHECR, this was first pointed out
in Ref.~\cite{chi}.  

The mean attenuation length of EHE $\gamma$-rays depends strongly on the
density of URB and on the strength of the EGMF, both of which are
uncertain at the present time. The EGMF typically
inhibits cascade development because of synchrotron cooling of the
$e^-e^+$ pairs produced in the PP process. Depending on the strength
of the EGMF, the synchrotron cooling time scale may be shorter than
the time scale of ICS, in which case the $e^-$ or the $e^+$ under
consideration loses energy through synchrotron radiation before it can
undergo ICS, and thus
cascade development stops. In this case the $\gamma$-ray flux is
determined mainly by the ``direct'' $\gamma$-rays, i.e., the ones that
originate at distances less than the absorption length due to PP
process. The energy lost by synchrotron cooling does not, however,
disappear --- rather, it appears at a lower energy and can 
initiate fresh EM cascades by interacting with the photons of a higher
energy background such as CMBR or IR/O depending on its energy. So the
overall effect
of a relatively strong EGMF is to deplete the $\gamma$-ray flux above some
energy in the EHE region and increase the $\gamma$-ray flux below a 
corresponding energy in the `low' (MeV -- GeV) energy region, 
where it will have to meet the constraints imposed by the measured
extragalactic diffuse $\gamma$-ray background~\cite{chen,sreekumar}. 

In addition to uncertainties in the strength of EGMF and the URB,
another major source of uncertainty in the cascade calculation is the 
poorly known IR/O background (see, e.g., Ref.~\cite{iro} for a recent
discussion of IR/O backgrounds). The latter strongly influences the  
cascade spectrum in the energy range from $\sim 10^{11}\ev$ to $\sim
10^{14}\ev$. Below this range, however, the cascade spectrum becomes
relatively insensitive~\cite{coppiahar} to the model parameters that
determine the IR/O background.   
This is a fortunate circumstance because this implies that the
constraints on TD models derived \cite{slc,slsc,ps,coppiahar} by comparing 
the measured 100 MeV -- 10 GeV $\gamma$-ray background with TD model
predictions are relatively insensitive to uncertainties in our precise
knowledge of the IR/O background, and hence are fairly robust. 
\subsubsection*{Neutrinos}
EHE neutrinos suffer absorption through 
fermion-antifermion pair production on the thermal background
neutrinos ($\nu + \bar{\nu}_b \to f \bar{f}$), where $f\equiv e, \mu,
\tau, \nu, q$, and $\nu_b$ is a thermal background neutrino.
Due to this absorption process, neutrinos of present observed 
energy $E_{\nu,0}$ would have to have been injected at redshifts less
than     
$z_a(E_{\nu,0})\simeq 3.5\times10^2 (10^{20}\ev/E_{\nu,0})^{2/7}$, for
$E_{\nu,0}\ga 3\times 10^{14}\ev$ \cite{bhs}. 

Actually, the $\mu$'s, $\tau$'s, and quarks created in the above
absorption process generate further neutrinos through decay of the
$\mu$'s and $\tau$'s and through decay of the charged pions created by
quark fragmentation. This leads to a ``neutrino cascade''
\cite{yoshida}, effectively increasing the size of the ``neutrino
horizon'' of the Universe. This, in turn, has the effect of increasing
the overall neutrino flux around $10^{20}\ev$ by a factor of few
relative to the case when the cascading effect is not taken into
account. 

Recently, it has been pointed out~\cite{weiler} that if neutrinos have a
small mass $m_\nu$ in the eV range, then EHE neutrinos (antineutrinos) of
energy $\sim 4\, (\ev/m_\nu)\times10^{12}\gev$ will annihilate on the relic
antineutrinos (neutrinos) to produce the Z-boson with a resonant cross
section of $\sim10^{-32}\cm^2$. The hadronic decay of the Z will produce
additional photons, neutrinos and nucleons, which will add to the photon
and neutrino cascading processes mentioned above. This process may also
leave specific signatures on the EHE neutrino spectrum, the detection of
which may in the end provide an indirect signature of the neutrino mass
and hence of dark matter. Detailed self-consistent calculation of
nucleon, photon and neutrino fluxes, however, remain to be done.     
\subsection*{Predicted Particle Fluxes and Constraints on TD
models}
As discussed above, the predicted particle fluxes depend on a number of
parameters: the X
particle mass $m_X$, the strength of EGMF, the URB, the injection spectra,
and so on. Recently, detailed numerical calculation of the particle
spectra, especially the spectrum of $\gamma$-rays in the entire energy
range from $\sim10^8\ev$ to $\sim10^{25}\ev$, have been done
~\cite{slc,slsc,ps} taking into account the effects of electromagnetic
cascading and EGMF. 
\begin{figure}[b!]
\centerline{\psfig{file=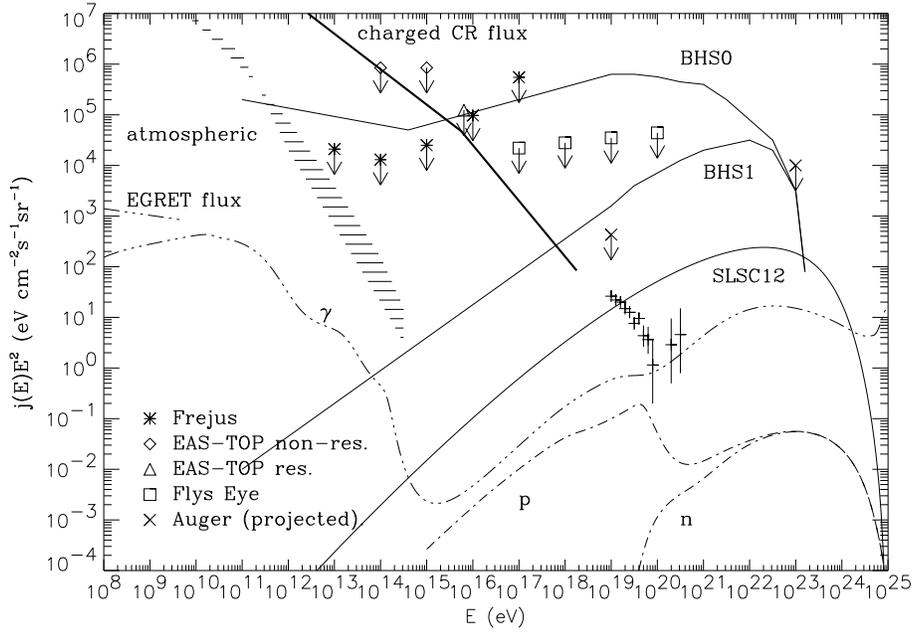,height=3.5in}}
\vspace{10pt}
\caption[...]{Predicted fluxes of $\gamma$-rays (dash - triple dotted
line), protons
and neutrons (dash-dotted lines) and $(\nu_\mu + \bar{\nu}_\mu)$ (solid
lines) for TD models with $p=1$, $m_X=10^{16}\gev$ and EGMF of $10^{-12}$
Gauss. The neutrino curve marked ``SLSC12'' corresponds to the calculation
of Ref.~\cite{slsc}, while the curves marked ``BHS0'' and ``BHS1''
represent the neutrino fluxes from Ref.~\cite{bhs} and correspond
respectively to $p=0$ and the $p=1$ models. Also shown are the estimates
of the atmospheric neutrino background for different zenith
angles~\cite{lipari} (hatched region marked ``atmospheric''). 
Data points with error
bars represent the combined cosmic ray
data from the Fly's Eye and AGASA experiments~\cite{hecrevents} above
$10^{19}\ev$, and the thick solid line represents piecewise power-law fit
to the observed charged CR flux. The dash - triple dotted line on
the left margin
represents experimental upper limits on the diffuse $\gamma$-ray flux at
100 MeV -- 5 GeV from EGRET data~\cite{chen,sreekumar}. Points with
arrows pointing downwards represent approximate upper limits on the
diffuse neutrino flux from Frejus~\cite{frejus}, the
EAS-TOP~\cite{eas-top}, and the Fly's Eye~\cite{fly'seye_nu} experiments,
as indicated. The projected limit shown for the proposed Auger experiment 
assumes the acceptance estimated in Ref.~\cite{auger_nu} for non-detection
over a five year period. (Courtesy G. Sigl)}
\label{F1}
\end{figure}
Fig.~1 shows the predicted diffuse particle fluxes for a representative
set of values of various parameters involved. 
Here I discuss only the predicted diffuse fluxes of particles assuming
uniform distribution of TD sources in the Universe. (Possible
individual ``bursting'' sources of EHECR and the possible reconstruction
of their ``images'' from data on  
arrival direction, arrival time and energy associated with individual 
events, leading to information
about the source characteristics and, in particular about the EGMF, are
discussed in the talk by G. Sigl; see this volume.) 

Because the magnitude of X particle production rate is not known {\it a
priori}, the best one can do at the present time is adopt a suitable 
normalization procedure for the absolute flux so as to be able to
explain the EHECR data, and then check whether
this normalization is consistent with all relevant observational data,
not just on EHECR particles, but also on other relevant particle
fluxes at lower energies, especially the diffuse gamma ray background
measurements in the MeV -- GeV region. The normalization of the absolute
fluxes in Fig.~1 is optimized in the sense that it has been determined by
fitting the `observable' particle (i.e., the combined nucleon and
$\gamma$-ray) fluxes to the measured EHECR data by the maximum likelihood
method~\cite{slsb} and corresponds to a likelihood significance of $>
50\%$. 

It is clear that TD models, while potentially
contributing dominantly to the particle fluxes above $\sim
10^{20}\ev$, make negligible contribution to cosmic ray flux below 
$\sim 10^{19}\ev$ because of the relatively hard nature of
the particle spectra in these models. Thus the flux below $\sim 2\times
10^{19}\ev$ is presumably due to a conventional acceleration
scenario, and was not included in the fitting procedure. 

Since pions are the most numerous particles in
the jets, their decay products, i.e., photons and neutrinos dominate the
number of particles at production. Since neutrinos suffer little 
attenuation and can come to us unattenuated from large cosmological
distances (except for absorption due to $e^+e^-$ pair production
by interacting with the cosmic thermal neutrino background, the path 
length for which is $\gg 100\mpc$), their fluxes, as expected, are the
largest among
all particles at the highest energies. However, their detection
probability is much lower compared
to those for protons and photons\footnote{The EHE neutrinos of TD origin 
would, however, be potentially detectable by the proposed 
space-based detectors like OWL and AIRWATCH, and ground based
detectors like Auger, Telescope Array, and so on; see articles on these
detector
projects in this volume.}. Photons also far outnumber nucleons at
production, and depending on the level of the
URB and EGMF, may dominate over the nucleon
flux and thus dominate the `observable' particle flux at EHECR energies. 

An important point to note is that photons and neutrinos in the TD
scenario are {\it primary} particles in that they are produced directly
from the decay of the pions in the hadronic jets. In contrast, photons and
neutrinos in conventional acceleration scenarios can be produced only
through {\it secondary} processes --- they are mainly produced by the
decay of photoproduced pions resulting from the GZK interactions of
primary HECR nucleons (produced by the acceleration process) with CMBR
photons~\cite{stecker_nu}. Of course, these
secondary neutrinos and photons would also be there in the TD
scenario, but their fluxes are sub-dominant to the primary
ones at the highest energies. 

The {\it shapes} of the EHE nucleon- and $\gamma$-ray spectra 
In the TD scenario are 
``universal'' \cite{bhs,abs} in the sense that they are 
independent of specific TD
process even though different TD sources evolve differently 
(as parametrized by the parameter $p$ in the X 
particle production rate). This is because, at these energies,
the attenuation lengths of nucleons and $\gamma$-rays are small ($\la
100\mpc$) compared to Hubble length and so the effects of
unknown cosmological evolution of the TD sources are 
negligible compared to propagation effects. The universal shapes of the
EHECR nucleon and $\gamma$-ray spectra reflect their injection spectra,
which, as discussed earlier, are determined by QCD. Thus, large
statistics measurement of the EHECR spectrum by future detectors may give
us a probe of new physics, and in particular QCD, at energies not
currently accessible in laboratory accelerators, {\it provided} a TD-like
non-acceleration scenario of origin of EHECR is correct.  

In contrast to the EHECR nucleons and $\gamma$-rays, the predicted
$\gamma$-ray flux below $\sim
10^{14}\ev$ (the threshold for pair production on the photons of CMBR)
and the predicted EHE neutrino flux depend on the total energy released
integrated over redshift and hence are dependent on the specific TD
model (i.e., specific value of $p$). 
In particular, the $\gamma$-ray flux below $\sim 10^{11}\ev$ 
scales as the total electromagnetic energy released from X particles
integrated over all redshifts and increases with decreasing
value of $p$. This has been used to constrain \cite{sigljedam} 
TD models from considerations of CMBR distortions and from independent
considerations of modified light element abundances due to ${}^4{\rm He}$
photodisintegration; for example, this rules
out \cite{sigljedam} the $p=0$ TD model. 

The $\gamma$-ray flux in Fig.~1 is consistent with the 
estimates of the upper limit on the background in the 100 MeV to $\sim$ 5
GeV from EGRET data~\cite{chen}. Very recently, 
new estimates of the background flux have been presented~\cite{sreekumar}
which now extend up to $\sim$ 100 GeV with roughly the same power-law
index as the earlier estimates~\cite{chen}. If the EGMF is 
significantly larger than the value assumed ($\sim 10^{-12}$ Gauss) for
the calculation of Fig.~1, then the electromagnetic cascade energy
transferred to the relevant low energy region will be higher, and in this
case the $\gamma$-ray flux in Fig.~1 will only be marginally
consistent, or may even be inconsistent, with the experimental diffuse
background flux estimates. Note, however, that 
the fluxes in Fig.~1 correspond to the case $m_X = 10^{16}\gev$. Lower
values of $m_X$ are possible, which will give lower contribution to the
low energy diffuse flux while at the same time producing enough energy in
the form of X particles to explain the observed EHECR flux. This can be
seen as follows: 

As we shall discuss in more details in the next section, 
for a relatively hard power-law photon injection spectrum $\propto
E^{-\alpha}$ with $\alpha < 2$, the energy injection rate $Q_0$ required
to match a given differential EHECR flux at a given energy decreases
with decreasing value of $m_X$ (provided, of course, $m_X > 10^{11}\gev$,
the energy of the highest energy event). For example, it is easy to see
(see next section) that for a photon injection spectrum $\propto
E^{-1.5}$, the value of $Q_0$ required to explain the EHECR flux is
roughly proportional to $m_X^{1/2}$. And since the cascade $\gamma$-ray
flux in the $\la$ 100 GeV region is essentially directly proportional to
$Q_0$ injected at EHECR energies, a reduction by a factor of 10, for
example, of the
predicted $\gamma$-ray flux in the $\la$ 100 GeV region is easily
achieved by reducing $m_X$ by about two orders of magnitude, i.e., for
$m_X\la10^{14}\gev$. And, of course, as already mentioned, the low energy
$\gamma$-ray flux is also reduced if the EGMF strength is lower. 

%
Thus, it seems that TD models of EHECR with
$m_X\sim10^{16}\gev$ and/or high EGMF ($\sim 10^{-9}$ Gauss) are
somewhat difficult
to simultaneously reconcile with EHECR data and the low energy diffuse
$\gamma$-ray background. On the other hand, models with
$m_X\la10^{14}\gev$ and low EGMF ($\la10^{-11}$ Gauss) can explain the
EHECR data and at the same time be consistent with
all existing data. 

The neutrino flux indicated by curve marked ``SLSC12'' in Figure
1 corresponds to overall X particle production rate obtained by the
maximum likelihood normalization of the `observable' (i.e., photon plus
nucleon) flux to the EHECR data for the $p=1$ TD model. The {\it
predicted} level of the neutrino flux, therefore, depends on the value
of EGMF, the radio background, etc., which go into determining the photon
flux. The earlier flux estimate (the curve marked ``BHS1'') in Fig.1 is
higher simply because the overall X particle production rate was
normalized to a higher value; that normalization was obtained 
by normalizing the predicted {\it proton} (as opposed to photon) flux
with the measured flux at a
lower energy $\sim5\times10^{19}\ev$ (where the measured cosmic ray flux
is higher than that at EHECR energies) --- the Fly's Eye and AGASA highest
energy events beyond $10^{20}\ev$ \cite{hecrevents} were
not yet discovered at that time! The $p=0$ model is ruled out by the upper
limits from Frejus as well as EAS-TOP experiments. (Actually, the $p=0$
model is also ruled out from considerations~\cite{sigljedam} of  
${}^4{\rm He}$ photo-disintegration and CMBR distortions as mentioned
above --- it
corresponds to unacceptably high rate of energy injection in the early
cosmological epochs.) On the other hand,
the new neutrino flux estimates for the $p=1$ model are consistent with
all existing experimental upper limits. At energies $\ga10^{20}\ev$,
the predicted neutrino flux in the $p=1$ TD model also dominates over the
predicted flux from blazars/AGNs as well as over the predicted flux
of ``cosmogenic''
neutrinos produced by interactions of UHE cosmic rays with CMBR
\cite{agn_cosmo_nu} (not shown in Fig. 1). For more details on the
detectability of neutrino fluxes in TD models, see, e.g.,
Ref.~\cite{slsc,yoshida,gandhi,blanco}. 

\section*{X-Particle Production Rate Required to Explain the EHECR Flux: 
A Benchmark Calculation}
To have an idea of the kind of numbers for the required X particle
production rate, we can perform a simple (albeit crude) benchmark
calculation as follows: 

Since in TD models, photons are expected to dominate the observable EHECR
flux, let us assume for simplicity that the highest energy events are due
to photons. Let us assume a typical three-body decay mode of the X into a
$q\bar{q}$ pair and a lepton: $X \to q\bar{q}\ell$. The two quarks will
produce two hadronic jets. Let $f_\pi$ denote the total pion fraction in a
jet. Then the photons from the two jets carry a total energy
$E_{\gamma,{\rm Total}}\simeq
\left({2\over3}\times0.9\times{1\over3}\right)m_X (f_\pi/0.9) = 0.2 m_X
(f_\pi/0.9)$. Let us assume a power-law hadronic fragmentation spectrum
with index 1.5. Then the photon injection spectrum from a single X
particle can be written as 
\beq
{dN_\gamma\over dE_\gamma} = {3\over m_X} \times 0.3 \, \left({3
E_\gamma\over m_X}\right)^{-1.5} \left({f_\pi\over
0.9}\right)\,,\label{photon_bench}
\eeq
which is properly normalized with the total photon energy $E_{\gamma,{\rm
Total}}$. We shall neglect cosmological evolution effects since photons of
EHECR energies have a cosmologically negligible path length of only $\sim$
few tens of Mpc for absorption through pair production on the radio
background. 

With these assumptions, the photon flux $j_\gamma (E_\gamma)$ at the
observed energy $E_\gamma$ is simply given by 
\beq
j_\gamma (E_\gamma)\simeq {1\over 4\pi} \lambda(E_\gamma)\, {dn_X\over
dt}\, {dN_\gamma\over dE_\gamma}\,,\label{flux_bench}
\eeq
where $\lambda(E_\gamma)$ is the pair production absorption path length of
a photon of energy $E_\gamma$. 
 
Noting that $dn_X/dt = Q_0/m_X$, and normalizing the above flux to the
measured EHECR flux corresponding to the highest energy event at $\sim
3\times10^{20}\ev$, given by $j(3\times10^{20}\ev)\approx
5.6\times10^{-41}\cm^{-2} \ev^{-1} \sec^{-1} \sr^{-1}$, we get 
\beq
Q_{0,{\rm required}}\approx 2.1\times10^{-21}\ev \cm^{-3} \sec^{-1}
\left({10\mpc\over\lambda_{\gamma,300}}\right)
\left({m_X\over10^{16}\gev}\right)^{1/2}\,,\label{Q_0_reqired}
\eeq
or 
\beq
\left({dn_X\over dt}\right)_{0,{\rm required}}\approx 2.1\times10^{-46}
\cm^{-3} \sec^{-1}
\left({10\mpc\over\lambda_{\gamma,300}}\right)
\left({m_X\over10^{16}\gev}\right)^{-1/2}\,,\label{Xrate_required}
\eeq
where $\lambda_{\gamma,300}$ is the absorption path length of a 300 EeV
photon (1 EeV $\equiv 10^{18}\ev$). The subscript 0 stands for the present
epoch. 

The above numbers are probably uncertain by up to an
order of magnitude depending on the exact form of the injection spectrum,
the absorption path length of EHECR photons, electromagnetic cascading
effect, and so on. Indeed, since electromagnetic cascading effect (which
we have neglected here) 
generally increases the photon flux, the above numbers are most likely 
overestimates. Nevertheless, they do serve as crude benchmark
numbers. 
These numbers indicate that in order for TDs to explain the EHECR events,
the X particles must be produced in the present epoch at a rate of $\sim
2\times 10^{35}\mpc^{-3}\yr^{-1}$, or in more ``down-to-earth'' units,
about $\sim 23 {\rm Au}^{-3}\yr^{-1}$, i.e., about 20 X particles within a
solar system-size volume per year. 

In the next section I discuss three plausible specific TD
processes and examine their efficacies with regard to X particle
production and EHECR keeping in mind the above rough estimates of the
required X particle production rate. 

\section*{X Particle Production from Cosmic Strings, Monopoles, and
Necklaces}
\subsection*{Cosmic Strings}
Let us first recall the salient features of evolution of cosmic strings
in the Universe; for a review, see Refs.~\cite{td}. Immediately after
their formation, the strings would be in a random tangled configuration.
One can define a characteristic length scale, $\xi_s$, of the initial
string configuration in terms of the overall mass-energy density,
$\rho_s$, of strings simply through the relation 
\beq
\rho_s = \mu/\xi_s^2\,,\label{string-density1}
\eeq
where $\mu$ denotes the string mass (energy) per unit length. 

Initially, the strings find themselves in a dense medium, so 
they move under a strong frictional damping force. The damping remains
significant until the temperature falls to $T\la (G\mu)^{1/2}\eta$, where
$G$ is Newton's constant and $\eta$ is the symmetry-breaking scale at
which strings were formed. [Recall, for GUT scale cosmic strings, for
example, $\eta\sim10^{16}\gev$, 
$\mu\sim \eta^2\sim(10^{16}\gev)^2$, and $G\mu\sim10^{-6}$.] In the
friction dominated epoch, a curved string segment of radius of curvature
$r$ quickly achieves a terminal velocity $\propto 1/r$. The small scale
irregularities on the strings are, therefore, quickly smoothed out. As a
result, the strings are straightened out and their total length shortened.   
The energy density in strings, therefore, decreases with
time. This means that the characteristic length scale $\xi_s$ describing
the string configuration increases with time as the Universe expands.
Eventually $\xi_s$ becomes comparable to the causal horizon distance
$\sim t$.  At about this time, the ambient density of the Universe 
becomes dilute enough that damping becomes unimportant so that
the strings start moving relativistically. 

Beyond this point, there are two possibilities.    
Causality prevents the length scale $\xi_s$ from growing faster than the
horizon length. So, either (a) $\xi_s$ keeps up with the horizon length,
i.e., $\xi_s/t$ becomes a constant, or (b) $\xi_s$ increases less rapidly
than $t$. 

Let us consider the second possibility first. In this case, the
string density falls less rapidly than $t^{-2}$. On the other
hand, we know that the radiation density in the radiation-dominated
epoch as well as matter density in the matter-dominated epoch both
scale as $t^{-2}$. Clearly, therefore, the strings would 
come to dominate the density of the Universe at some point of time. 
It can be shown that this
would happen quite early in the history of the Universe unless the
strings are very light, much lighter than the GUT scale strings. A string
dominated early Universe would be unacceptably inhomogeneous conflicting
with the observed Universe\footnote{However, a string dominated {\it 
recent} Universe --- dominated by `light' strings formed at a phase
transition at about the electroweak symmetry breaking scale --- is
possible. Such a string dominated recent Universe may even have some
desirable cosmological properties~\cite{string-dom}. Such light strings
are, however, not of interest to us in this discussion.}.       

The other possibility, which goes by the name of ``scaling'' hypothesis,
seems to be more probable, as suggested by detailed numerical
as well as analytical studies~\cite{td,vincent}. The numerical simulations
generally find that the string density does reach the scaling regime given
by $\rho_{s,{\rm scaling}}\propto 1/t^2$, and then continues to be in this
regime. It is, however, clear that in order for this to happen, strings
must lose energy in some form at a certain rate. This is because, in
absence of any energy loss, the string configuration would only be
conformally stretched by the expansion of the Universe on scales larger
than the horizon so that $\xi_s$ would only scale as the scale factor
$\propto t^{1/2}$ in the radiation dominated Universe, and $\propto
t^{2/3}$ in the matter dominated Universe. In both cases, this would fail
to keep the string density in the scaling regime, leading back to
string domination. In order for the string density
to be maintained in the scaling regime, energy must be lost by the string 
configuration per unit proper volume at a rate
$\dot{\rho}_{s,{\rm loss}}$ satisfying the equation 
\beq
\dot{\rho}_{s,{\rm total}} = -2 {\dot{R}\over R} \rho_s + 
\dot{\rho}_{s,{\rm loss}}\,,\label{loss}
\eeq
where the first term on the right hand side is due to expansion of the
Universe, $R$ being the scale factor of the expanding Universe. In
the scaling regime $\dot{\rho}_{s,{\rm total}} = -2 \rho_s/t$, which gives 
$\dot{\rho}_{s,{\rm loss}} = - \rho_s/t$ in the radiation dominated
Universe, and $\dot{\rho}_{s,{\rm loss}} = -(2/3) \rho_s/t$ in the matter
dominated Universe. 

The important question is, in what form does the string configuration lose
its energy so as to maintain itself in the scaling regime? One possible
mechanism of energy loss from strings is 
formation of closed loops. Occasionally, a segment of 
string might self-intersect by curling up on itself. The
intersecting segments may intercommute, i.e., ``exchange partners'' ,
leading to formation of a closed loop which pinches off the string. 
The closed loop would then oscillate and lose energy by emitting
gravitational radiation and eventually disappear. It can be shown that
this is indeed an efficient mechanism of extracting energy from 
strings and transferring it to other forms. The string energy loss rate
estimated above indicates that scaling could be maintained by roughly of
order one closed loop of horizon size ($\sim t$) formed in a horizon size
volume ($\sim t^3$) in one hubble expansion time ($\sim t$) at any time
$t$. In principle, as far as energetics is concerned, one can have the
same effect if, instead of one or few large loops, a large number of
smaller loops are formed. Which one may actually happen depends on the
detailed dynamics of string evolution, and can only be decided by means of
numerical simulations.         

Early numerical simulations seemed to support the large (i.e., $\sim$
horizon size) loop formation
picture. Subsequent simulations with improved resolution, however, found a
lot of small-scale
structure on strings, the latter presumably being due to kinks left on the
strings after each crossing and intercommuting of string segments.
Consequently, loops formed were found to be much smaller in size than
horizon size and correspondingly larger in number. Further simulations
showed that the loops tended to be formed predominantly on the scale of
the cut-off length imposed for reasonable resolution of even the smallest
size loops allowed by the given resolution scale of the simulation. 
It is, however, generally thought
that the small-scale structure cannot continue to build up indefinitely,
because the back-reaction of the kinky string's own gravitational field
would
eventually stabilize the small-scale structure at a scale of order 
$G\mu t$. The loops would,
therefore, be expected to be formed predominantly of size $\sim G\mu t$,
at any time $t$. Although much smaller than the horizon size, these loops
would still be of `macroscopic' size, much larger than the microscopic
string width scale ($\sim\eta^{-1}\sim\mu^{-1/2}$). These loops would,
therefore, also oscillate and eventually disappear by emitting
gravitational radiation. Thus, according to above picture, the dominant
mechanism of energy loss from strings responsible for maintaining the
string density in the scaling regime would be formation of 
macroscopic-size ($\gg \eta^{-1}$) loops and emission of  
gravitational radiation by these loops. More details of this picture of
cosmic string evolution can be found in Refs.~\cite{td}. 

How is the above picture of cosmic string evolution relevant for cosmic
rays? How does one get X particles from cosmic strings? 

One way of getting X particles from cosmic strings is through the
so-called cusp-evaporation mechanism~\cite{cusp}. I will not discuss this
mechanism here, but the resulting X particle production rate
generally turns out to be too low to be of relevance to EHECR.   

Another possibility arises as follows: As the closed loops oscillate and
emit gravitational radiation, they lose energy and shrink. Eventually,
when a loop's radius shrinks to a size of order the width of the string,
the string unwinds and turns into X particles, which will then decay, 
producing high
energy particles. However, each loop in the end only produces of the order
of one X particle. The resulting cosmic ray flux is again too low to be
observable~\cite{gk}. 

Clearly, the only way cosmic strings may produce large number of X
particles is if macroscopic lengths of string are involved in the X
particle production process. One such mechanism was suggested in
Refs.~\cite{pbkofu,br} based on the following arguments. The picture of
gravitational radiation being the dominant energy loss mechanism for 
cosmic strings rests on the
assumption that the loops themselves do not self-intersect frequently.
Since the
motion of a freely oscillating loop is periodic (with a period of 
$L/2\, $, $L$ being the invariant length of the loop \cite{kibble-turok}),
a loop formed 
in a self-intersecting configuration will undergo self-intersection within
its first period of oscillation. If the loop does indeed self-intersect
and break up into two smaller loops, and if the daughter loops again
self-intersect breaking up into two even smaller loops, and so on, then
one can see that a single initially large loop of size $L$ can break up
into a debris of tiny loops of size $\eta^{-1}$ (thereby turning
into X particles) within a time scale of $\sim L$. Since the
largest loops are expected to be of size $< t$, one sees that the above
time scale can be much smaller than one hubble time. In other words, one
large loop can break up into a large number of tiny loops (X particles)
within one hubble time. Such self-intersecting loops would not 
radiate much energy gravitationally because that would require many
periods of oscillation. In other words, such repeatedly 
self-intersecting loops would be a channel through which energy
contained in macroscopic lengths of string could go into X particles
instead of into gravitational radiation. As shown in 
Ref.~\cite{kibble-turok}, some non-circular loops could also be in
configurations which would collapse completely into double-line
configurations and subsequently annihilate into X particles~\cite{bkt}. 
It has also been argued in Ref.~\cite{siemens} that the self-intersection
probability of a loop increases exponentially with the number of
small scale kinks on a loop. 

At the time of work of Ref.~\cite{br}, however, it was not clear as to
what fraction of the string energy might go into X particles through 
processes discussed above. Treating 
this fraction to be a free parameter $f$, i.e., assuming that a fraction
$f$ of the energy extracted from the long strings per unit volume per unit
time went into X particles, the authors of
Ref.~\cite{br} found that the fraction $f$ had to be rather small,
$f\la7\times10^{-4}$; otherwise, the predicted cosmic ray flux from
GUT scale cosmic strings would exceed the observed flux. 
Results of more recent calculation of the predicted observable particle
flux~\cite{slsc} (see Fig. 1) correspond to an upper
limit on $f$ which is about two orders of magnitude lower\footnote{This is
due to the fact that the `observable' particle flux now includes
the gamma ray flux in addition to the protons ---
in contrast to only the much lower proton flux considered in
Ref.~\cite{br} --- and
also because of the maximum likelihood normalization of the predicted flux to
the observed data.}. 

The actual value of $f$ is still unknown. But if  
gravitational radiation, and not massive particle radiation, is the
dominant energy loss mechanism for cosmic strings, then the fraction $f$
may actually be much smaller than the above upper limit, in which case the 
flux of cosmic rays produced by cosmic strings through X particles would
be small and below the observed flux. However, rapid conversion of
a significant fraction of the energy in macroscopically large loops into
tiny loops (X particles) through repeated loop self-intersections due, for
example, to the presence of large number of kinks on the
loops~\cite{siemens}, and consequent production of significant EHECR flux, 
cannot be ruled out. 

Very recently, the notion gravitational radiation as the dominant
energy loss mechanism for cosmic strings has been
questioned by the results of new numerical simulations of
evolution of cosmic strings~\cite{vincent}. Authors of Ref.~\cite{vincent}
claim that if loop production is not artificially restricted by imposing a
cutoff length for loop size in the simulation, then loops tend to be
produced
dominantly on the smallest allowed length scale in the problem, namely, on
the scale of the width of the string. Such small loops promptly collapse 
into X particles. In other words, there is essentially
no loop production at all --- the string energy density is maintained in
the scaling regime by energy loss from strings predominantly in the form
of direct X particle emission, rather than by formation of large loops
and their subsequent gravitational radiation. This new result, while
subject to confirmation by independent simulations, obviously has
important implications for EHECR. Indeed, in this case, the upper limit on
the fraction $f$ mentioned above implies severe
constraint on GUT-scale cosmic strings. From the results of
Ref.~\cite{vincent}, the string energy density in the scaling regime is
$\rho_{s,{\rm scaling}}\simeq \mu/(0.3\, t)^2$. With X particle production
as the dominant energy loss mechanism, we immediately see from
Eq.~(\ref{loss}) that rate of production of X particles from strings per
unit volume must be $dn_X/dt\simeq 7.4 (\mu/m_X) t^{-3}$ in the matter
dominated
era. Taking, for cosmic strings, $\mu\simeq \pi\eta^2$, and taking
$m_X\sim 0.7\eta$, where $\eta$ is the symmetry-breaking scale, we get
by using the constraint imposed by Eq.~(\ref{Xrate_required}),
$\eta\la4.2\times10^{13}\gev$. Thus the GUT scale cosmic strings with
$\eta\sim10^{16}\gev$ are ruled out by EHECR data if the
results of Ref.~\cite{vincent} are correct. At the
same time X particles from cosmic strings formed at a phase transition
with $\eta\sim 10^{13}$ -- $10^{14}\gev$ are able to explain the EHECR
data. Cosmic strings may thus be a ``natural'' source of extremely high
energy cosmic rays if massive particle radiation, and {\it not}
gravitational radiation, is indeed their dominant energy loss mechanism. 

Cosmic string formation with $\eta\sim10^{14}\gev$ rather than at the GUT
scale of $\sim 10^{16}\gev$ is not hard to envisage. For example, the
symmetry breaking scheme SO(10) $\to$ SU(3) $\times$ SU(2) $\times$
U$(1)_{\rm Y} \times$ U(1) can take place at the GUT unification scale
$M_{\rm GUT}\sim10^{16}\gev$; with no U(1) subgroup broken, this phase
transition produces no strings. However, the second U(1) can be
subsequently broken with a second phase transition at a scale $\sim
10^{14}\gev$ to yield the cosmic strings relevant for EHECR. Note that
these strings would be too light to be relevant for structure formation in
the Universe and their signature on the CMBR sky would also be too weak to
be detectable. Instead, the extremely high energy end of the cosmic ray
spectrum may offer a probing ground for signatures of these cosmic
strings. 
\subsection*{Monopoles}
If monopoles were formed at a phase transition in the early Universe,
then, as Hill~\cite{hill} suggested in 1983, a metastable 
monopole-antimonopole bound state --- ``monopolonium'' --- is possible. 
At any temperature $T$, monopolonia would be formed with binding 
energy $E_b\ga T$. The initial radius $r_i$ of a monopolonium would be 
$r_i\sim {1\over2} g_m^2/E_b$, where $g_m$ is the magnetic charge
(which is related to the electronic charge $e$ through the Dirac
quantization condition $eg_m=N/2$, $N$ being the monopole's winding
number). Classically, of course, the monopolonium is unstable. Quantum
mechanically, the monopolonium can exist only in certain ``stationary''
states characterized by the principal quantum number $n$ given by $r=n^2
a_m^{\rm B}$, where $n$ is a positive integer, $r$ is the instantaneous
radius, and $a_m^{\rm B}=8\alpha_e/m_M$ is the ``magnetic'' Bohr radius of
the monopolonium. Here $\alpha_e=1/137$ is the ``electric'' fine-structure
constant, and $m_M$ is the mass of a monopole. Since the Bohr
radius of a monopolonium is much less than the Compton wavelength (size) 
of a monopole, i.e., $a_m^{\rm B}\ll m_M^{-1}$, the
monopolonium does not exist in the ground ($n=1$) state, because the
monopole and the antimonopole would be overlapping, and so would
annihilate each other. However, a monopolonium would initially be formed
with $n\gg1$; it would then undergo a series of transitions through a
series of tighter and tighter bound states by emitting initially photons
and subsequently gluons, Z bosons, and finally the GUT X bosons.
Eventually, the cores of the monopole and the antimonopole would overlap, 
at which point the monopolonium would annihilate into X particles. Hill
showed that the life time of a monopolonium is proportional to the cube of
its initial radius. Depending on the epoch of formation, some of the
monopolonia formed in the early Universe could be surviving in the
Universe today, and some would have collapsed in the recent epochs. 
It can be shown~\cite{bs} that monopolonia collapsing in the present
epoch would have been formed in the early Universe at around the epoch of
primordial nucleosynthesis. 

The X particles produced by the collapsing monopolonia may give rise to
EHECR. This possibility was studied in details in Ref.~\cite{bs}, who
showed that this process, like cosmic strings, can also be described by
an equation for X particle production rate described by Eq.~(\ref{xrate})
with $p=1$. The efficacy of the process, however, depends on two
parameters, namely, (a) the monopolonium-to-monopole fraction at formation 
($\xi_f$) and (b) the monopole abundance. The latter is unknown, while
the former is in principle
calculable by using the classical Saha ionization formalism. 
However, phenomenologically, since a monopole mass can be typically
$m_M\sim 40 m_X$ (so that each monopolonium collapse can release $\sim$ 
80 X particles), we see from Eq.~(\ref{Xrate_required}) that one
requires roughly (only!) about 3 monopolonium collapse per decade
within a volume roughly of the size of the solar system. Whether or not
this can
happen depends, as already mentioned, on $\xi_f$ as well as on the
monopole abundance, the condition~\cite{bs} being $(\Omega_M
h_{100}^2) h_{100}\xi_f \simeq 1.7\times10^{-8} (m_X/10^{16}\gev)^{1/2}
(10\mpc/\lambda_{\gamma,300})$, where $\Omega_M$ is the mass density
contributed by monopoles in units of closure density of the Universe,
and $h_{100}$ is the present hubble constant in units of $100 \km
\sec^{-1}
\mpc^{-1}$. Thus, as expected, larger the monopole abundance, smaller is
the monopolonium fraction $\xi_f$ required to explain the EHECR flux.
Note that, since $\xi_f$ must always be less than
unity, the above requirements can be satisfied as long as 
$(\Omega_M h_{100}^2) h_{100} > 1.7\times10^{-8} (m_X/10^{16}\gev)^{1/2}$.  
Recall, in this context, that 
the most stringent bound on the monopole abundance is given by the Parkar
bound~\cite{kt}, $(\Omega_M h_{100}^2)_{\rm Parkar}\la 4\times10^{-3}
(m_M/10^{16}\gev)^2$. The estimate of $\xi_f$ obtained by using the Saha
ionization formalism~\cite{hill,bs} shows that the resulting requirement
on the monopole abundance (in order to explain the EHECR flux) is
well within the Parkar bound mentioned above. The monopolonium collapse, 
therefore, seems to be an attractive scenario in this regard. 
A detailed numerical simulation of monopolonium formation to determine
the monopolonium fraction at formation would, however, be useful in
this context.  
\subsection*{Necklaces}
A cosmic necklace is a possible hybrid topological defect consisting of
a closed cosmic string loop with monopole ``beads'' on them. Such a hybrid
defect was first considered by Hindmarsh and 
Kibble~\cite{beads}. Such defects could be formed in a two
stage symmetry-breaking scheme such as $G \to H \times U(1) \to H
\times Z_2$, where $Z_2$ is the discrete group \{-1,1\} under
multiplication. In such a symmetry breaking, monopoles are formed at the
first step of the symmetry breaking if the group $G$ is semisimple. In the
second step, the so-called ``${\rm
Z}_2$'' strings are formed and then each monopole gets attached to two
strings, with the monopole magnetic flux channeled along the strings. 
Possible production of massive X particles from necklaces has been pointed
out in Ref.~\cite{necklace}. 

The evolution of the necklace system is not well understood. The crucial
quantity is the dimensionless ratio $r\equiv m_M/(\mu d)$, where $m_M$
denotes the monopole mass, $\mu$ is the string mass per unit length, and
$d$ is the average separation between the monopoles. For $r\ll 1$, the
monopoles play a subdominant role, and the evolution of the system is
similar to that of ordinary cosmic strings. For $r\gg 1$, on the other
hand, the monopoles determine the behavior of the system. The monopoles
sitting on the strings tend to make the motion of the closed necklaces
aperiodic. The authors of Ref.~\cite{necklace} assume that closed
necklaces undergo frequent self-intersections, leading to monopole -
antimonopole annihilation and, consequently, release of massive X
particles. The X particle production rate for the
necklace system can also be written in the form of Eq.~(\ref{xrate}) with
$p=1$. The efficacy of the process depends on free
parameters, $r$, $\mu$ and $m_M$. For appropriate choice of the
parameters, the required EHECR flux can be obtained. For more details see
Ref.~\cite{necklace} and the article by Berezinsky in this volume. 

\subsection*{Topological Defects in Supersymmmetric Theories}
Recently, it has been pointed out~\cite{susytd} that in a wide class of
supersymmetric unified theories, the higgs bosons associated with the
gauge symmetry breaking can be `light' --- of mass of order the soft
supersymmetry breaking scale $\sim$ TeV --- even though the gauge
symmetry breaking scale (and hence the mass of the gauge boson) itself may
be much
larger. The topological defects in these theories can, therefore,
simultaneously be sources of $\sim$ TeV mass-scale higgs bosons as well as
supermassive (mass up to $\sim 10^{16}\gev$) gauge bosons. The decay of
the TeV higgs may give a significant contribution to the observed diffuse
$\gamma$-ray background above a few GeV, while the supermassive gauge
boson decay may explain the EHECR. For more details, see
Ref.~\cite{susytd}.  
\section*{Conclusion}
There is no dearth of specific models of X particle producing processes
involving topological defects. Almost all the ``realistic'' processes
studied so far can be parametrized in the form of Eq.~(\ref{xrate}) with
$p=1$. The spectra of various kinds of particles produced for all these
processes would essentially be similar to the ones shown in Fig.~1.
Different processes might, however, contribute different amounts to the
total flux. It is difficult to say which particular process may contribute
most to the observed EHECR flux. However, in this respect, the cosmic
string scenario seems to be relatively parameter free, especially if the
new results of Ref.~\cite{vincent} are correct. Experimentally, it will be
difficult, if not impossible, to tell which specific TD process, if at
all, is responsible for the EHE cosmic rays. The best one can hope for is
that proposed future experiments like OWL, Auger, and so on, may be able   
to tell us whether topological defects (or, for that matter, any {\it
non-acceleration} mechanism in general) or some other completely
different scenario is involved in producing the observed EHECR.  
In any event, the prospect of being able to look for signatures of new
physics with the proposed EHECR experiments is certainly exciting, to say
the least.  

{\it Note added}: Recently it has been pointed out~\cite{tdbau} that if
indeed decays
of X particles from TD processes are responsible for the EHECR, then the
same TD processes occurring in the early Universe may also have given  
rise to
the observed baryon asymmetry of the Universe through CP- and Baryon
number violating decays of the X particles. 

\section*{Acknowledgments}
I am grateful to G\"unter Sigl for many helpful discussions, and in
particular for providing the Figure. This work is supported by a NAS/NRC
Resident Senior Research Associateship at NASA/GSFC. 


\begin{thebibliography}{}

\bibitem{kibble} T.W.B. Kibble, {\it J. Phys.} {\bf A9}, 1387 (1976).  

\bibitem{td} A. Vilenkin and E.P.S. Shellard, {\it Cosmic Strings and
other Topological Defects} (Cambridge Univ. Press, Cambridge, 1994); 
M.~Hindmarsh and T.~W.~B.~Kibble, Rep. Prog. Phys. {\bf 58}, 477 (1995);
T.~W.~B.~Kibble, {\it Aust. J. Phys.} {\bf 50}, 697 (1997); T.~Vachaspati,
{\it Topological Defects in the Cosmos and Lab} (hep-ph/9802311) (to
appear in {\it Contemporary Physics}. 

\bibitem{bhs} P. Bhattacharjee, C.T. Hill, and D.N. Schramm, {\it Phys.
Rev. Lett.} {\bf 69}, 567 (1992). 

\bibitem{hill} C.T. Hill, {\it Nucl. Phys.} {\bf B 224}, 469 (1983). 

\bibitem{hsw} C.T. Hill, D.N. Schramm, and T.P. Walker, {\it Phys. Rev.}  
{\bf D 36}, 1007 (1987). 

\bibitem{cusp} J.~H.~MacGibbon and R.~H.~Brandenberger, {\it Nucl. Phys.} 
{\bf B 331}, 153 (1990); P. Bhattacharjee, {\it Phys. Rev.} {\bf D 40},
3968 (1989);  M.~Mohazzab and R.~Brandenberger, {\it Int. Jour. Mod.
Phys.} {\bf D 2}, 183 (1993). 

\bibitem{pbkofu} P. Bhattacharjee, in {\it Astrophysical Aspects of
the Most Energetic Cosmic Rays}, eds. M.~Nagano and F.~Takahara (World
Scientific, Singapore, 1991), {\it pp}. 382 -- 399. 
 
\bibitem{br} P. Bhattacharjee and N.C. Rana, {\it Phys. Lett.} 
{\bf B 246}, 365 (1990). 

\bibitem{gk} A.~J.~Gill and T.~W.~B.~Kibble, {\it Phys. Rev.} {\bf D 50},
3660 (1994).  

\bibitem{bs} P.~Bhattacharjee and G.~Sigl, {\it Phys. Rev.} {\bf D 51},
4079 (1995). 

\bibitem{necklace} V. Berezinsky and A. Vilenkin, {\it Phys. Rev. Lett.}
{\bf 79}, 5202 (1997). 

\bibitem{hecrevents} D.J. Bird {\it et al}, {\it Phys. Rev. Lett.} 
{\bf 71}, 3401 (1993); {\it Astrophys. J.} {\bf 441}, 144 (1995);
N.~Hayashida {\it et al}, {\it Phys. Rev. Lett.} {\bf 73}, 3491 (1994);
S.~Yoshida {\it et al}, {\it Astropart. Phys.} {\bf 3}, 105 (1995). 

\bibitem{ssb} G. Sigl, D.N. Schramm, and P. Bhattacharjee, {\it Astropart.
Phys.} {\bf 2}, 401 (1994). 

\bibitem{slsb} G.~Sigl, S.~Lee, D.~N.~ Schramm, and P. ~Bhattacharjee,
{\it Science}, {\bf 270}, 1977 (1995).  

\bibitem{slc} G. Sigl, S. Lee, and P. Coppi, {\it astro-ph/9604093}. 

\bibitem{slsc} G.~Sigl, S.~Lee, D.~N.~ Schramm, and P.~Coppi, {\it Phys.
Lett.} {\bf B 392}, 129 (1997). 

\bibitem{ps} R.J. Protheroe and T. Stanev, {\it Phys. Rev. Lett.} {\bf
77}, 3708 (1996); {\bf 78}, 3420 (1997) (E). 

\bibitem{hillas} A.~M.~Hillas, {\it Ann. Rev. Astron. Astrophys.} {\bf
22}, 425 (1984). 

\bibitem{norman} C.A. Norman, D.B. Melrose, and A. Achterberg, {\it
Astrophys. J.} {\bf 454}, 60 (1995). 

\bibitem{drury} See, e.g., L. O'C Drury, {\it Rep. Prog. Phys.} {\bf 46},
973 (1983); F.~C.~Jones and D.~C.~Ellison, {\it Sp. Sci. Rev.} {\bf 58},
259 (1991). 

\bibitem{biermann} J.P. Rachen and P.L. Biermann, {\it Astron. Astrophys.}
{\bf 272}, 161 (1993); P.~L.~Biermann, in this volume. 

\bibitem{es} J.~W.~Elbert, and P.~Sommers, {\it Astrophys. J.}
{\bf 441}, 151 (1995).

\bibitem{pbicnapp} P. Bhattacharjee, in {\it Non-Accelerator Particle
Physics}, Ed. R.~Cowsik (World Scientific, Singapore, 1995). 

\bibitem{pbtokyo} P. Bhattacharjee, in {\it Extremely High Energy Cosmic
Rays: Astrophysics and Future Observatories}, Ed. M.~Nagano (ICRR, Univ.
of Tokyo, 1996). 

\bibitem{siglheidel} G. Sigl, {\it Sp. Sci. Rev.} {\bf 75}, 375 (1996).  

\bibitem{abs} F.A.~Aharonian, P.~Bhattacharjee, and D.N.~Schramm,
{\it Phys. Rev.} {\bf D 46}, 4188 (1992).  
 
\bibitem{yoshida} S.~Yoshida, {\it Astropart. Phys.} {\bf 2}, 187 (1994); 
S.~Yoshida, H.~Dai, C.~Jui, and P.~Sommers, {\it Astrophys. J.} {\bf 479},
547 (1997). 

\bibitem{chi} X.~Chi, C.~Dahanayake, J.~Wdowczyk, and
A.~W.~Wolfen\-dale, {\it Astropart. Phys.} {\bf 1}, 129 (1993); 
{\it ibid}. {\bf 1}, 239 (1993). 

\bibitem{kt} E.W. Kolb and M.S. Turner, {\it The Early Universe}
(Addison-Wesley, Redwood City, California, 1990). 

\bibitem{helium3} C. B\"auerle {\it et al}, {\it Nature} {\bf 382}, 332
(1996); V.~Ruutu {\it et al}, {\it Nature} {\bf 382}, 334 (1996). 

\bibitem{zurek} W.H. Zurek, {\it Nature} {\bf 317}, 505 (1985). 

\bibitem{witten} E. Witten, {\it Nucl. Phys.} {\bf B 249}, 557 (1985). 

\bibitem{vincent} G. Vincent, N.D. Antunes, and M. Hindmarsh, {\it Phys.
Rev. Lett.} (to be published) (hep-ph/9708427); G.~R.~Vincent,
M.~Hindmarsh, and M.~Sakellariadou, {\it Phys. Rev.} {\bf D 56}, 637
(1997). 

\bibitem{gzk} K.~Greisen, {\it Phys. Rev. Lett.} {\bf 16}, 748 (1966);
G.~T.~Zatsepin and V.~A.~Kuzmin, {\it Pisma Zh. Eksp. Teor. Fiz.} 
{\bf 4}, 114 (1966) [{\it JETP. Lett.} {\bf 4}, 78 (1966)]; F.~W.~Stecker,
{\it Phys. Rev. Lett.} {\bf 21}, 1016 (1968); F.~A.~Aharonian and 
J.~W.~Cronin, {\it Phys. Rev.} {\bf D 50}, 1892 (1994). 

\bibitem{lund} B. Andersson {\it et al}, {\it Phys. Rep.} {\bf 97}, 31
(1983). 

\bibitem{cluster} G. Marchesini and B.R. Webber, {\it Nucl. Phys.} {\bf B
310}, 461 (1988); G.~Marchesini {\it et al}, {\it Comp. Phys. Comm.} {\bf
67}, 465 (1992); HERWIG version 5.9, hep-ph/9607393.  

\bibitem{jetset} T. Sj\"ostrand and M. Bengtsson, {\it Comp. Phys. Comm.}
{\bf 43}, 367 (1987). 

\bibitem{ariadne} L. L\"onnblad, {\it Comp. Phys. Comm.} {\bf 71}, 15
(1992). 

\bibitem{lphd} Ya. I. Azimov, Yu.~L.~Dokshitzer, V.~A.~Khoze, and
S.~I.~Troyan, {\it Z. Phys.} {\bf C 27}, 65 (1985); {\bf C 31}, 213
(1986). 

\bibitem{lphdreview} Yu. L. Dokshitzer, V.~A.~Khoze, A.~H.~Mueller, and
S.~I.~Troyan, {\it Basics of perturbative QCD} (Editions Frontiers,
Saclay, 1991); R.~K.~Ellis, W.~J.~Stirling, and B.~R.~Webber, {\it QCD and
Collider Physics} (Cambridge Univ. Press, Cambridge, England, 1996); 
V.~A.~Khoze and W.~Ochs, {\it Int. J. Mod. Phys.} (To be published)
(hep-ph/9701421). 

\bibitem{mlla} A.H. Mueller, {\it Nucl. Phys.} {\bf B213}, 85 (1983); {\it
ibid.} {\bf B241}, 141 (1984) (E). 

\bibitem{stecker_nu} F.W. Stecker, {\it Astrophys. J.} {\bf 228}, 919
(1979). 

\bibitem{chen} A.~Chen, J.~Dwyer, and P.~Kaaret, {\it Astrophys. J.} 
{\bf 463}, 169 (1996). 

\bibitem{sreekumar} P. Sreekumar {\it et al}, {\it Astrophys. J.} {\bf
494} (1998) (in press) (astro-ph/9709257); 
P.~Sreekumar, F.~W.~Stecker, and S.~C.~Kappadath, astro-ph/9709258 (in 
press). 

\bibitem{iro} M.H. Salamon and F.W. Stecker, {\it Astrophys. J.} (1998)
(in press); astro-ph/9704166.  

\bibitem{coppiahar} P.~S.~Coppi and F.~A.~Aharonian, {\it Astrophys. J.}
{\bf 487}, L9 (1997). 

\bibitem{weiler} T.J. Weiler, {\it E-print} hep-ph/9710431. 

\bibitem{lipari} P. Lipari, {\it Astropart. Phys.} {\bf 1}, 195 (1993). 

\bibitem{frejus} W.~Rhode {\it et al.}, {\it Astropart. Phys.} 
{\bf 4}, 217 (1996). 

\bibitem{eas-top} M. Aglietta {\it et al}, in {\it Proc. 24th Int. Cosmic
Ray Conf.} {\bf 1}, 638 (1995). 

\bibitem{fly'seye_nu} R.M. Baltrusaitis {\it et al}, {\it Astrophys. J.}
{\bf 281}, L9 (1984); {\it Phys. Rev.} {\bf D 31}, 2192 (1985). 

\bibitem{auger_nu} G. Parente and E. Zas, {\it E-print} astro-ph/9606091. 

\bibitem{sigljedam} G.~Sigl, K.~Jedamzik, D.~N.~Schramm, and V.~Berezinsky,
{\it Phys. Rev.} {\bf D 52}, 6682 (1995). 

\bibitem{agn_cosmo_nu} F.W. Stecker, C. Done, M.~H.~Salamon, and
P.~Sommers, {\it Phys. Rev. Lett.} {\bf 66}, 2697 (1991); {\bf 69}, 2738
(1992) (E). 

\bibitem{gandhi} R. Gandhi, C. Quigg, M.H. Reno, and I.~Sarcevic, {\it
Astropart. Phys.} {\bf 5}, 81 (1995). 

\bibitem{blanco} J.J. Blanco-Pillado, R.A. Vazquez, and E. Zas, {\it Phys.
Rev. Lett.} {\bf 78}, 3614 (1997). 

\bibitem{string-dom} A. Vilenkin, {\it Phys. Rev. Lett.} {\bf 53}, 1016
(1984); D.~Spergel and Ue-Li Pen, {\it Astrophys. J.} {\bf 491}, L67
(1997). 

\bibitem{kibble-turok} T.W.B. Kibble and N. Turok, {\it Phys. Lett.} {\bf
B 116}, 141 (1982). 

\bibitem{bkt} P. Bhattacharjee, T.W.B. Kibble and N. Turok, {\it Phys.
Lett.} {\bf B 119}, 95 (1982). 

\bibitem{siemens} X.A. Siemens and T.W.B. Kibble, {\it Nucl. Phys.} {\bf B
438}, 307 (1995).  

\bibitem{beads} M.B. Hindmarsh and T.W.B. Kibble, {\it Phys. Rev. Lett.}
{\bf 55}, 2398 (1985).

\bibitem{susytd} P. Bhattacharjee, Q. Shafi, and F.W. Stecker, {\it 
E-print} hep-ph/9710533. 

\bibitem{tdbau} P. Bhattacharjee, {\it E-print} hep-ph/9803223. 

\end{thebibliography}
\end{document}